\documentclass[twocolumn,prl,showpacs,preprintnumbers,amsmath,amssymb]{revtex4-1}
\usepackage{graphicx}
\usepackage{dcolumn}
\usepackage{bm}
\usepackage{tabularx}
\usepackage{multirow}

\begin{document}

\title { Graphene-like Dirac states and Quantum Spin Hall Insulators in the square-octagonal $MX_2$ ($M$=Mo, W; $X$=S, Se, Te) Isomers}

\author{Yan Sun$^1$}
\author{Claudia Felser$^1$}
\author{Binghai Yan$^{1,2}$}
\email[Corresponding author:]{yan@cpfs.mpg.de}

\affiliation{$^1$ Max Planck Institute for Chemical Physics of Solids, 01187 Dresden, Germany}
\affiliation{$^2$ Max Planck Institute for the Physics of Complex Systems, 01187 Dresden,Germany}

\date{\today}

\begin{abstract}

We studied the square-octagonal lattice of the transition metal dichalcogenide MX$_2$ (with $M$=Mo, W; $X$=S, Se and Te), as an isomer of the normal hexagonal compound of MX$_2$. 
By band structure calculations, we observe the graphene-like Dirac band structure in 
a rectangular lattice of MX$_2$ with nonsymmorphic space group symmetry. 
Two bands with van Hove singularity points cross each at the Fermi energy, 
leading to two Dirac cones that locates at opposite momenta. 
Spin-orbit coupling can open a nontrivial gap at these Dirac points and induce the quantum spin Hall (QSH) phase, the 2D topological insulator. 
Here, square-octagonal MX$_2$ structures realize the interesting graphene physics, such as Dirac bands and QSH effect, 
in the transition metal dichalcogenides.

\end{abstract}

\pacs{73.20.At, 71.20.-b, 71.70.Ej}

\maketitle

Since the discovery of graphene\cite{graphene1,graphene2},
research about two dimensional (2D) materials has been widely
explored in both theory and experiment. During
the the last few years, transition metal dichalcogenides (TMD) $MX_2$
(with $M$=Mo, W and $X$=S, Se, Te)~\cite{MoS1, MoS2} have attracted
extensive attention.  
For example,  monolayer of MoS$_2$ is a direct-gap semiconductor and regarded as massive Dirac systems 
in the honeycomb lattice with interesting valley physics~\cite{MoS3}. 
Very recently an isomer structure of $MX_2$ in the square-octagonal (so)
lattice~\cite{Li} was found to exhibit gapless band structure with a Dirac cone a the zone center, 
and lattice distortion~\cite{Terrones} was claimed to remove above Dirac cone and induce additional band crossing
at the Fermi energy. 
However, the topological feature in the band structure was neglected due to the missing of spin-orbital coupling in calculations. 
In this work, we revisited the square-octahedral lattices of $MX_2$ isomers, 
and discovered their graphene-like Dirac band structures (see Fig. 1) and the 2D topological insulator phase
~\cite{kane3, shoucheng2}.

Inspired by the atomic structures of  grain boundaries in normal hexagonal MoS$_2$ ~\cite{Najmaei,Zande,Zhang}, 
 a square-octagon (so) lattice o  for MoS$_2$~\cite{Li} was investigated in theory.
 There are four Mo atoms and eight S atoms in each primitive cell,
as shown in Fig.1. And this lattice can be viewed as repeated
square-octagon pairs in both $x$ and $y$ directions. 
Here, the trigonal prismatic structure of MoS6 is slightly distorted compared to the hexagonal phase.
Further, distortions from the square lattice to a rectangle lattice (distorted-so lattice) 
were found to optimize the strain in the 2D structure and realize more stable structures~\cite{Terrones}. 
In this work, we adopted the similar lattice structures to TMD $MX_2$ (with $M$=Mo and W; $X$= S, Se and Te) monolayers and found that the distorted-so-lattice is indeed the energetically favored structure for all compounds. The lattice distortion and total energy differences
are summarized in Table I. We can see that for lattice distortion, characterized by the ratio of in-plane lattice parameters $b/a$, becomes stronger as the increasing of atomic
radius of atom $X$ for fixed metal atom $M$,  due to the elongated $M$-$X$ bond.

\begin{figure}[htbp]
\begin{center}
\includegraphics[width=0.5\textwidth]{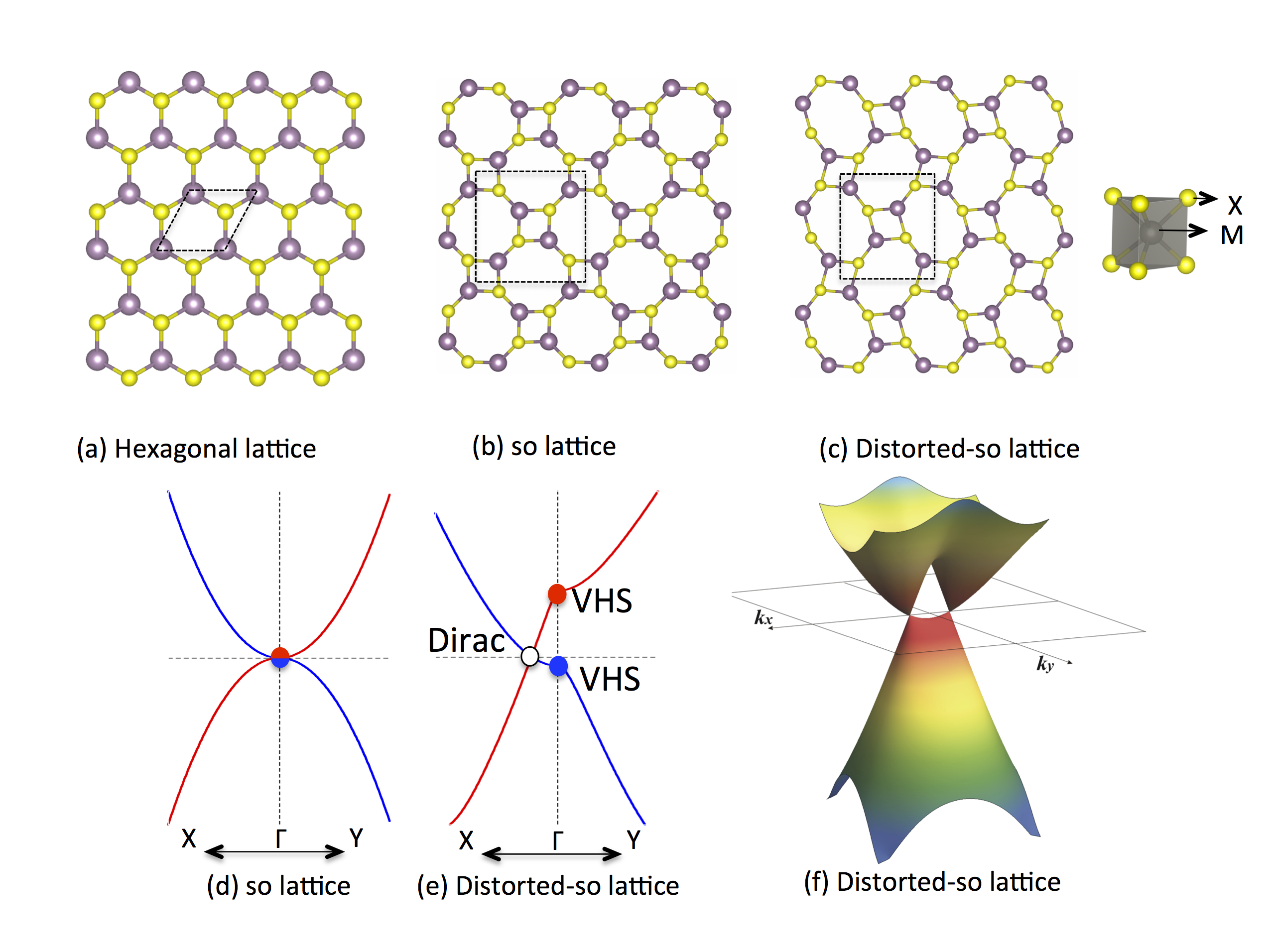}
\end{center}
\caption{
(color online) Crystal lattice structure for $MX_{2}$ (M = Mo, W; X = S, Se, Te) in the
(a) normal hexagonal lattice, (b) square-octagonal (so) lattice and (c) distorted-so lattice.
The primitive unit cell is denoted by dashed lines. The band structures near the Fermi energy are illustrated in the 
2D Brillouin zone for (d) so lattice, (e) distorted-so lattice. The 3D plot of the distorted-so lattice is shown in (f), where
two Dirac cones forms in the $\Gamma - X$ line. The Dirac point and the van Hove singularity (VHS) point are indicated by 
empty and filled circles in (e), respectively.
}
\label{crystal}
\end{figure}

\begin{table*}
\caption{ 
The lattice constants and electronic properties of $MX_2$
(with $M$=Mo, W; $X$=S, Se and Te). Total energy differences between
two types of lattices is defined as
$\triangle E=E_{so-lattice}-E_{distorted-so-lattice}$
in one primitive cell. In the distorted-so-lattice, the linear band
crossing at Fermi energy induces a Dirac cone,
which locates on the line of $\Gamma$-$X$ in the lattice momentum space.
Its detailed location ($k_D$) is defined as the the relative distance away
from $\Gamma$ point, as the schematic diagram given in Fig.1(b). The
$ab- initio$ calculations for lattice structure optimization, total energy
differences  $\triangle E$, and location of Dirac cone are preformed
without the inclusion of SOC, while for band gap $E_g$ and $Z_2$ invariants
the SOC was included.
}

\begin{tabular}{|l|l|l|l|l|l|l|l|l|l|l|l|}
\hline
\multicolumn{1}{|c|} {} & \multicolumn{2} {|c|} {lattice constant (\AA)}  & \multicolumn{1}{|c|}{$\triangle E$ (meV)} &\multicolumn{2} {|c|}{$E_g$ (meV)}  & \multicolumn{2} {|c|} {$Z_2$ invariant} & \multicolumn{1} {|c|} {$k_D$ $(\frac{2\pi}{a})$ } \\
\hline
\multicolumn{1}{|c|}{} & \multicolumn{1} {|c|}{so-lattice} & \multicolumn{1} {|c|}{distorted-so-lattice} & \multicolumn{1}{|c|}{} & \multicolumn{1} {|c|}{so-lattice} & \multicolumn{1} {|c|}{distorted-so-lattice} & \multicolumn{1} {|c|}{so-lattice} & \multicolumn{1} {|c|}{distorted-so-lattice} & \multicolumn{1} {|c|}{} \\
\hline
\multicolumn{1}{|c|}{MoS$_2$} & \multicolumn{1} {|c|}{$a$=6.34} & \multicolumn{1} {|c|}{$a$=6.30, $b$=6.37}  & \multicolumn{1}{|c|}{0.68} & \multicolumn{1} {|c|}{25} & \multicolumn{1} {|c|}{12}   & \multicolumn{1} {|c|}{1} & \multicolumn{1} {|c|} {1} & \multicolumn{1} {|c|}{0.025} \\
\hline
\multicolumn{1}{|c|}{MoSe$_2$} & \multicolumn{1} {|c|}{$a$=6.62} & \multicolumn{1} {|c|}{$a$=6.56, $b$=6.67}  & \multicolumn{1}{|c|}{1.11} & \multicolumn{1} {|c|}{38} & \multicolumn{1} {|c|}{23} & \multicolumn{1} {|c|}{1} & \multicolumn{1} {|c|} {1}  & \multicolumn{1} {|c|}{0.073} \\
\hline
\multicolumn{1}{|c|}{MoTe$_2$} & \multicolumn{1} {|c|}{$a$=7.06} & \multicolumn{1} {|c|}{$a$=6.72, $b$=7.32}  & \multicolumn{1}{|c|}{17.87} & \multicolumn{1} {|c|}{49} & \multicolumn{1} {|c|}{19} & \multicolumn{1} {|c|}{1} & \multicolumn{1} {|c|} {1} & \multicolumn{1} {|c|}{0.201} \\
\hline
\multicolumn{1}{|c|}{WS$_2$}  & \multicolumn{1} {|c|}{$a$=6.36} & \multicolumn{1} {|c|}{$a$=6.32, $b$=6.42}  & \multicolumn{1}{|c|}{1.26} & \multicolumn{1} {|c|}{111} & \multicolumn{1} {|c|}{64}  & \multicolumn{1} {|c|}{1} & \multicolumn{1} {|c|} {1} & \multicolumn{1} {|c|}{0.072} \\
\hline
\multicolumn{1}{|c|}{WSe$_2$} & \multicolumn{1} {|c|}{$a$=6.64} & \multicolumn{1} {|c|}{$a$=6.30, $b$=6.88} & \multicolumn{1}{|c|}{29.50} & \multicolumn{1} {|c|}{152} & \multicolumn{1} {|c|}{20}  & \multicolumn{1} {|c|}{1} & \multicolumn{1} {|c|} {1}  & \multicolumn{1} {|c|}{0.225} \\
\hline
\multicolumn{1}{|c|}{WTe$_2$}  & \multicolumn{1} {|c|}{$a$=7.11} & \multicolumn{1} {|c|}{$a$=6.66, $b$=7.37} & \multicolumn{1}{|c|}{87.20} & \multicolumn{1} {|c|}{213} & \multicolumn{1} {|c|}{19}  & \multicolumn{1} {|c|}{1} & \multicolumn{1} {|c|} {1} & \multicolumn{1} {|c|}{0.245} \\
\hline

\end{tabular}\label{tab_energy}
\end{table*}

We mainly focus on their topological
electronic properties, after clarifying the lattice structures of $MX_2$.
Our calculations  have been performed
by using density functional theory (DFT) with projected augmented
wave (PAW) method as implemented in the code of Vienna $Ab$ $initio$
Simulation Package (VASP)~\cite{Kresse, Kresse2}. The exchange-correlation
energy are considered in the generalized gradient approximation (GGA)
level with Perdew-Burke-Ernzerhof (PBE) based density functional~\cite{pbe}.
The energy cutoff was set to be 350 eV. The tight binding matrix elements
were calculated by projection Bloch states onto maximally localized
Wannier functions (MLWFs)~\cite{marzari,souza,mostofi}, using the
VASP2WANNIER90 interface~\cite{Franchini}.

Because all the compounds of $MX_2$ share similar electronic properties,
in the following part we will take MoS$_2$ and WS$_2$ as the examples for
detailed analysis of their electronic structures. Band structures for
two types of lattices are compared in Fig. 2. Without the inclusion of
SOC effect, band structure in the so-lattice
presents doubly degenerated $d_{z2}$ states at $\Gamma$ point around
the Fermi energy for both MoS$_2$ and WS$_2$, exhibiting as a
semi-metallic state. As long as the
SOC is taken into consideration, the degenerated $d_{z2}$ bands 
split into two single states, with one locating at the top of valence
band and the other up shifting to the second conduction band. Beside
that, band anti-crossing between conduction and valence bands appears
around $\Gamma$ point, which implies the existence of band inversion.

In order to make clear the topological electronic properties, we analyzed
the wave functions around the Fermi energy. As presented in Fig.2(a) and (d),
the top of valence band and bottom of conduction bands at $\Gamma$ point
are mainly dominated by $M-d_{z2}$ and $M-d_{x2-y2}$ orbitals, respectively,
and these two states have opposite parities. Since the splitted two $M-d_{z2}$
states are both with plus parities, the effect of SOC here is just opening
the band gap but not changing topological band order. Because of the inversion
symmetry in the so-lattice, we can directly achieve the topological number of
$Z_2$ invariant by the products of parities at time reversal invariant
momentas (TRIMs)~\cite{Fu}. The parity products for occupied states at the
three independent TRIMs of $\Gamma$(0,0),
$X\left((0.5,0),(0,0.5)\right)$  and
$S\left(0.5,0.5\right)$ are -, + and +, respectively,
which gives the $Z_2$ invariant $\nu_0=1$. Therefore, it directly conforms the
existence of QSH insulator state in the so-$MX_{2}$.

\begin{figure}[htbp]
\begin{center}
\includegraphics[width=0.45\textwidth]{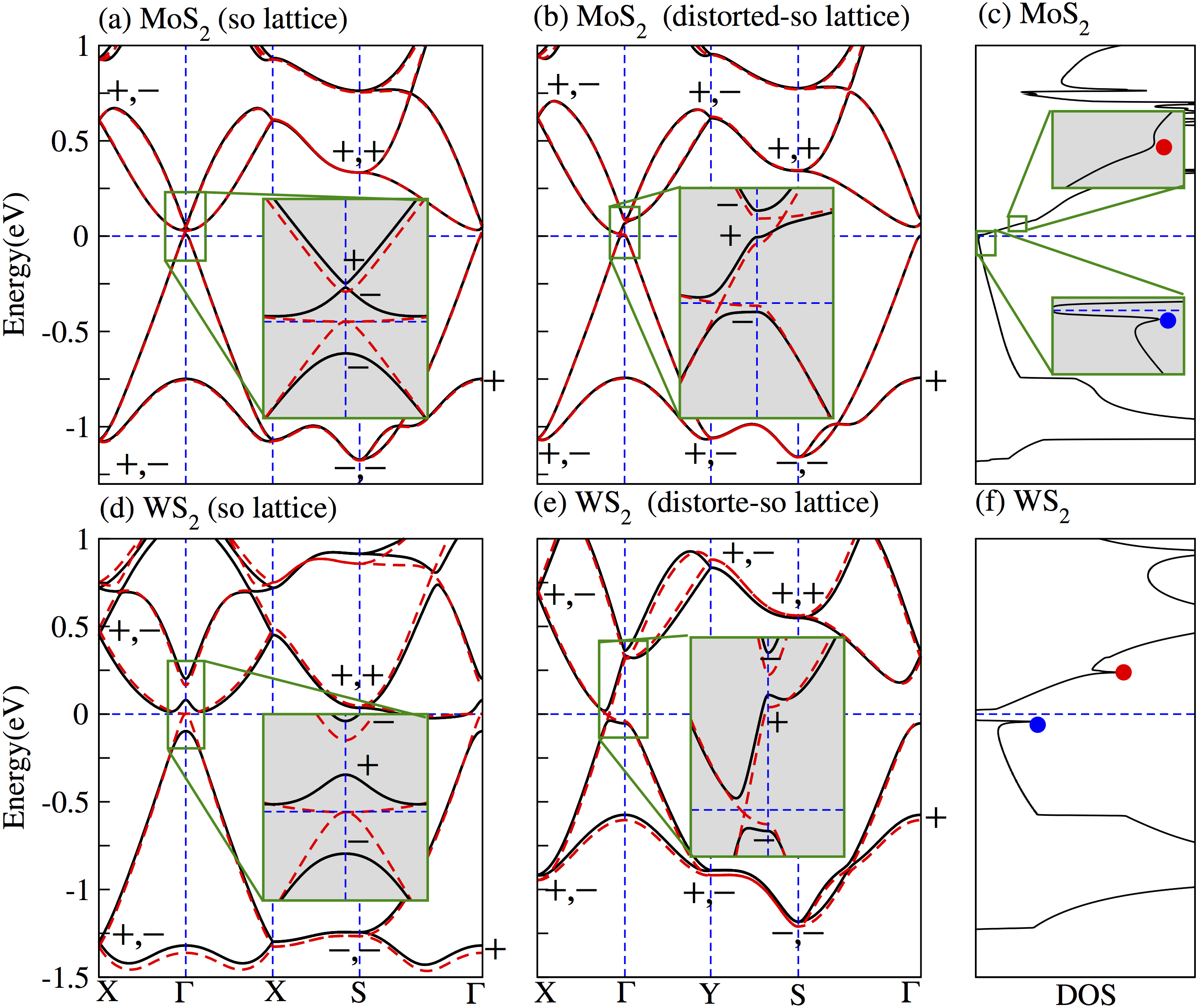}
\end{center}
\caption{
(color online) Electronic band structures and density of states (DOS) for 
MoS$_2$ and WS$_2$. Energy dispersion along high symmetry lines in 2D 
BZ for (a, b) MoS$_2$ with so and distorted-so lattices, and (d, e) WS$_2$ 
with so and distorted-so lattices. (c) and (f) are total DOS for distorted-so
MoS$_2$ and WS$_2$, respectively. Energy bands with (black solid lines) and 
without (red dot lines) are both included. DOS are calculated with the inclusion
of SOC Local band structures around 
Fermi energy are also given in the inserted figures. VHS points in the 
DOS are remarked with red and blue dots. Plus and minus signs remarked in 
band structures are the parity eigenvalues. Band structures are calculated 
from $ab-initio$ method, and DOS are calculated from MLWFs. Fermi energy 
is set to be zero.
}
\label{band_structure}
\end{figure}

After the lattice distortion, the symmetry decreased from
$D_{2h}$ to $C_{2h}$, and one can expect a dramatically
change of the electronic properties. As shown in Fig.2(b)
and (e), doubly degenerated $M-d_{z2}$ states at $\Gamma$
point split for both MoS$_2$ and WS$_2$, even without the inclusion
of SOC. Meanwhile, a linear band crossing forms a massless Dirac cone
at the Fermi level near $\Gamma$ point on the line of $\Gamma$-$X$.
The crossed two bands are mainly dominated by $M-d_{z2}$ and
$M-d_{x2-y2}$ orbitals, respectively. Besides, it is found that
the location of the 2D Dirac cone $k_D$ is strongly related
to the distortion strength. As given in Table I, the Dirac cone
shifts far away from $\Gamma$ point as the increasing of lattice
distortion. With the inclusion of SOC, as presented by the local
band structures in Fig.2(b) and (e), a general gap is
opened with the breaking of the band crossing, which
is just a typical image for TIs. Since the distortion does
not change the bulk band order, electronic structures are
topological equivalent in so-lattice and distorted-so-lattice.
For further confirmation, we also calculated the $Z_2$ invariant.
Though the distortion changes the lattice structure and atomic
positions, inversion symmetry is preserved. So parity product at
TRIMs is still effective for identifying the topological
order~\cite{Fu}. Our calculations found that the $Z_2$ invariant
is 1 for any compound of $MX_2$, as shown in Table I. Therefore,
distorted-so-$MX_2$ is still locating at topological non-trivial state.

The SOC effect becoming stronger along with the increasing
of atomic weight, and correspondingly, the SOC opened band gap
should be also increased. As given in Table I, it is really the
case in so-lattice. However, for the distorted-so-lattice, the
lattice distortion is also becoming stronger as the increasing
of atomic radius. Meanwhile, the band crossing point $k_D$
shifts far away from $\Gamma$ point, as shown in Table I.
Hence, the bulk band gap in the
situation with distorted-so-lattice is decided by a competition
between the strengths of SOC and lattice distortion. As presented
in Table I, the competed result in this series of compounds gives
the largest band gap of about 64 meV, appearance in WS$_2$.

Because of the non-trivial 2D bulk band order in $MX_2$,
topological protected metallic edge state happens.
For calculation of the edge state, we have constructed
the slab model through MLWFs based tight binding
method~\cite{marzari,souza,mostofi}. In so-lattice, since the bands
around Fermi level are almost consisted by the hybridized $M-d_{z2}$
and $M-d_{x2+y2}$ orbitals for all the compounds of $MX_{2}$,
MLWFs are derived from atomic $d_{z2}$ and $d_{x2-y2}$-like orbitals.
While for distorted-so-lattice, due to the difference between lattice
constants of $a$ and $b$, it is not accurate enough to describe the
tight binding model by only including $d_{z2}$ and $d_{x2-y2}$-like
orbitals, in which $d_{xy}$-like oribtal is also necessary. The tight binding
parameters are determined from the MLWFs overlap matrix. The slab model
was constructed in $x$ direction for so-lattice due to the cubic symmetry,
and both of $x$ and $y$ directions were choose for distorted-so lattice,
corresponding projected 1D-BZ given in Fig.1 (e). In order to eliminate the
coupling between two edges, the widenesses of the slabs were up to
200 and 100 unit cells for MoS$_2$ and WS$_2$, respectively.

From Fig.3 we can see that edge states exist for both so-lattice and
distorted-so-lattice. While the details are depending on different
compounds and edge terminations. For example, edge bands cut Fermi level
three times for so-WS$_2$, whereas other cases just cut Fermi level once.
As the differences of chemical potentials, some cases do not show Dirac
point like edge states, but the non-trivial  $Z_2$ invariant guarantees
the edge bands always cutting Fermi level odd times.

\begin{figure}[htbp]
\begin{center}
\includegraphics[width=0.45\textwidth]{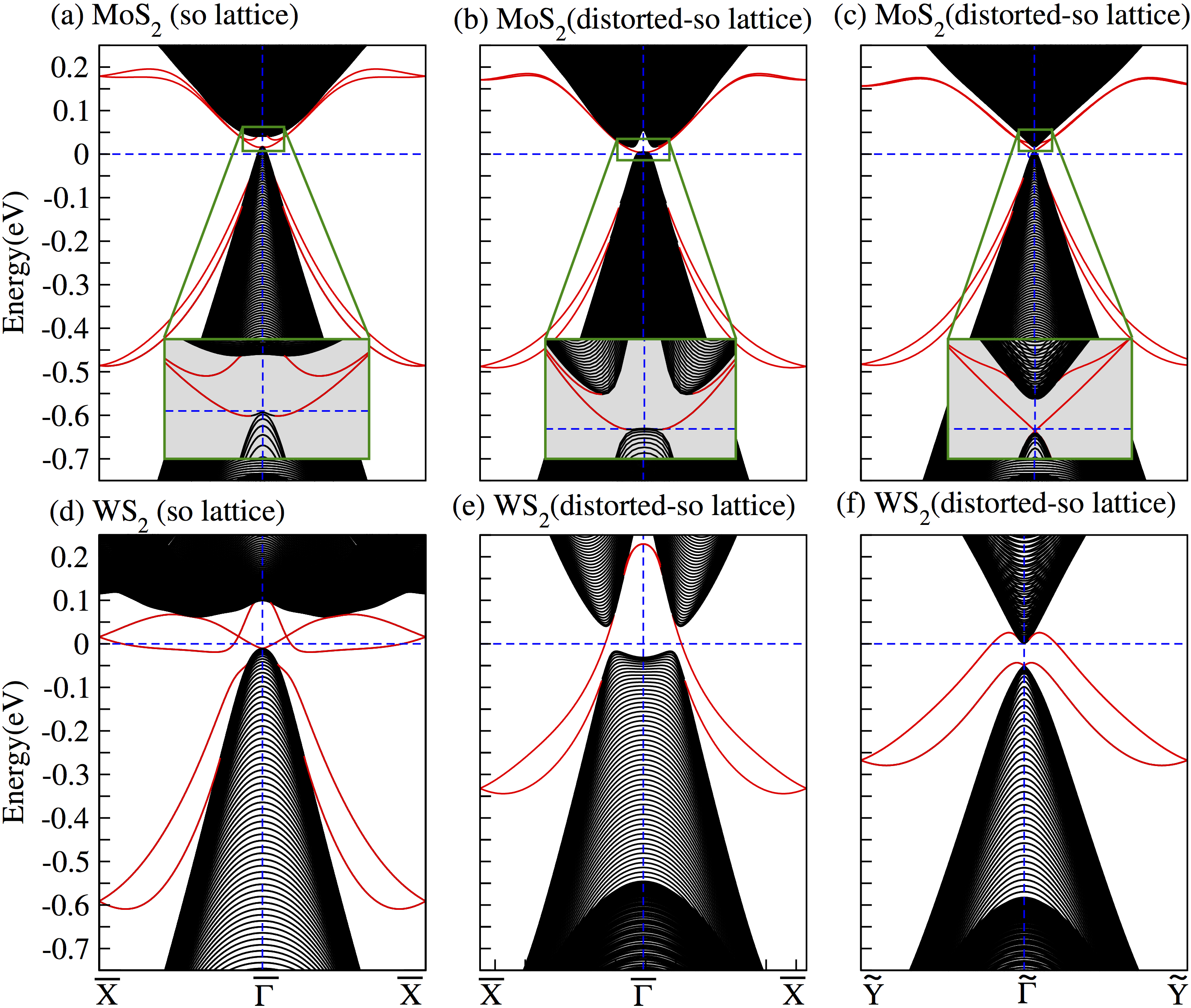}
\end{center}
\caption{
(color online) Tight binding edge band structures for (a) so-MoS$_2$,
(b) distorted-so-MoS$_2$, (c) so-WS$_2$,as well as (d) distorted-so-MoS$_2$.
Local band structures for MoS$_2$ around Fermi energy are showing inset.
Red curves corresponding to edge states. Fermi energy is set to zero.
}
\label{dege}
\end{figure}


As we have discussed above, the effect of SOC is just opening
the band gap but not inducing band inversion. Therefore, the
physics of band inversion can be understood without SOC. As
in the situation of graphene, the band inversion in $MX_2$ is
due to the symmetry protected Van Hove singularity (VHS).
In two dimensions, energy band with opposite energy dispersions
along different directions can induce the saddle point at the
transition momenta, which known as VHS. Moreover, if two bands
with VHS cross each other, a band inversion will be there
naturally. This is just case in the $MX_2$.

From the DFT band structures we have known that, without inclusion
of SOC the so-lattice-$MX_2$ presents as semi-metallic state with
degenerated $M-d_{z2}$ locating at $\Gamma$ point. This degeneration
can be regarded as an overlap of two saddle points belonging the two
bands with VHSs, as the schematic showing in Fig. 1(d).
After lattice distortion, the charactor of VHS becomes more obviously
due to the symmetry breaking. As presented in Fig. 1(e), the two
neighboring bands shifts  upwardly and downwardly, respectively, and
overlapped two saddle points are separated by a band gap around
$\Gamma$ point. The features of VHS are also obviously in the local
DFT band structures as given in Figs. 2 (a, b) and (d, e). Besides,
according to the density of states (DOS) expression in lattice momentum
space,
$g(E)\propto\int\frac{dk_{i}}{(\nabla_{k_{j}}E)}$ (where, $g(E)$ is the
DOS at energy $E$, and $i,j$=$x,y$), DOS integrand at the saddle point
diverges due to the  extreme of energy dispersion.  The diverged peaks can
be found in the DOS for both MoS$_2$ and WS$_2$, as presented in Fig2.(c)
and (f).

Furthermore, as we have seen in the DFT band structures, the lattice
distortion also shifts the degenerated point away
from $\Gamma$ point. Similar to graphene, the new degenerated point
exhibits as a massless Dirac cone, and the only difference is the location
difference of the Dirac cone in lattice momentum space.
In graphene, the Dirac cones paired locate at the high symmetry
momentas of $K$ and $K'$, which are connected by inversion and time
reversal symmetry. While in distorted-so-$MX_2$, Dirac cones paired locate
at $k_D$ on the line of $X$-$\Gamma$-$X$,
which are also connected by inversion and time reversal symmetry, as shown
in Fig. 4(b) and (c). SOC is strong enough
to open a considerable bang gap in distorted-so-$MX_{2}$, 
different from the case of graphene with very weak SOC~\cite{kane1}.

In conclusion, by first principles calculations we have theoretically proposed a series of QSH insulators
in the allotropes of monolayer TMD $MX_{2}$ (with $M$=Mo, W; $X$=S, Se and Te ). 
The ground states in the allotropes show distorted-so-lattice.
Similar to graphene, the band inversion is induced
by symmetry protected VHS, and SOC just plays the role of opening band gap. 
However, due to much stronger SOC effect in $MX_2$,  the inverted band gaps are big enough for measurement. 
The band gap is about 12 meV in MoS$_2$. And the largest band gap in this series QSH
insulators is about 64 meV, which appears in in WS$_2$. The QSH 
phase in $MX_{2}$ broaden the physical properties for
DMTCs. As the so and distorted-so
phase are derived from the grain boundary of  h-MoS$_2$ monolayers, it is 
possible to experimentally detect the QSH phase in MoS$_2$ and the other TMD
materials through measuring the current or photo emission on their 
boundary. In addition, the VHS points indicate the existence of Lifshitz transition in the Fermi surface~\cite{Lifshitz} 
and might promise interesting superconductivity in these materials.

\begin{acknowledgments}
We are grateful to Z. Wang, C.-X. Liu, H. Su for fruitful discussion.
\end{acknowledgments}

\end{document}